\documentclass[a4paper,11pt]{article}\usepackage{jcappub}\def\PRDstyle#1{}\def\JCAPstyle#1{#1}\let\Abstract\abstract
\usepackage[utf8]{inputenc}
\usepackage[T1]{fontenc}
\usepackage{cmap}

\def\imo{i}
\def\re#1{Re(#1)}
\def\im#1{Im(#1)}
\def\K{{\cal K}}

\begin{document}

\title{Quasinormal ringing of regular black holes in asymptotically safe gravity: the importance of overtones}

\JCAPstyle{
\author[1]{R.~A.~Konoplya,}\emailAdd{roman.konoplya@gmail.com}
\author[1]{A.~F.~Zinhailo,}\emailAdd{antonina.zinhailo@physics.slu.cz}
\author[2]{J.~Kunz,}\emailAdd{jutta.kunz@uni-oldenburg.de}
\author[1]{Z.~Stuchlík,}\emailAdd{zdenek.stuchlik@physics.slu.cz}
\author[3]{A.~Zhidenko}\emailAdd{olexandr.zhydenko@ufabc.edu.br}
\affiliation[1]{Research Centre for Theoretical Physics and Astrophysics, \\ Institute of Physics, Silesian University in Opava, \\ Bezručovo náměstí 13, CZ-74601 Opava, Czech Republic}
\affiliation[2]{Department of Mathematics and Natural Sciences, University Oldenburg,\\ Carl-von-Ossietzky-Str. 9-11, D-26129 Oldenburg, Germany}
\affiliation[3]{Centro de Matemática, Computação e Cognição (CMCC), \\ Universidade Federal do ABC (UFABC), \\ Rua Abolição, CEP: 09210-180, Santo André, SP, Brazil}
\arxivnumber{2206.14714}
}

\PRDstyle{
\author{R. A. Konoplya}\email{roman.konoplya@gmail.com}
\affiliation{Research Centre for Theoretical Physics and Astrophysics, Institute of Physics, Silesian University in Opava, Bezručovo náměstí 13, CZ-74601 Opava, Czech Republic}
\author{A. F. Zinhailo}\email{antonina.zinhailo@physics.slu.cz}
\affiliation{Research Centre for Theoretical Physics and Astrophysics, Institute of Physics, Silesian University in Opava, Bezručovo náměstí 13, CZ-74601 Opava, Czech Republic}
\author{J. Kunz}\email{jutta.kunz@uni-oldenburg.de}
\affiliation{Department of Mathematics and Natural Sciences, University Oldenburg, Carl-von-Ossietzky-Str. 9-11, D-26129 Oldenburg, Germany}
\author{Z. Stuchlík}\email{zdenek.stuchlik@physics.slu.cz}
\affiliation{Research Centre for Theoretical Physics and Astrophysics, Institute of Physics, Silesian University in Opava, Bezručovo náměstí 13, CZ-74601 Opava, Czech Republic}
\author{A. Zhidenko} \email{olexandr.zhydenko@ufabc.edu.br}
\affiliation{Centro de Matemática, Computação e Cognição (CMCC), Universidade Federal do ABC (UFABC), \\ Rua Abolição, CEP: 09210-180, Santo André, SP, Brazil}
}

\Abstract{
Asymptotically safe gravity is based on the idea that the main contribution to the Schwarzschild-like black hole spacetime is due to the value of the gravitational coupling which depends on the distance from the origin and approaches its classical value in the far zone. However, at some stage this approach has an arbitrariness of choice of some identification parameter. The two cases of identification are considered here: first, by the modified proper length (the Bonanno-Reuter metric), and second, by the Kretschmann scalar (the metric for this case coincides, up to the redefinition of constants, with the Hayward metric). Even though the quasinormal modes of these metrics have been extensively studied, a number of interesting points were missed. We have found that quasinormal modes are qualitatively similar for both types of identification. The deviation of the fundamental mode from its Schwarzschild limit may be a few times larger than it was claimed in the previous studies. The striking deviation from the Schwarzschild limit occurs for overtones, being as large as hundreds of percent even when the fundamental mode is almost coinciding with the Schwarzschild one. This happens because the above metrics are very close to the Schwarzschild one everywhere, except a small region near the event horizon, which is crucial for overtones. The spectrum of both metrics contains purely imaginary (non-oscillatory) modes, which, for some values of parameters, can appear already at the second overtone.
}

\maketitle

\section{Introduction}\label{Intro}
Construction of the model of a quantum corrected black hole is important not only as a particular strong-gravity solution within, yet an unknown, non-contradicting quantum gravity, but also because of the fundamental problem related to the final stage of Hawking evaporation and central singularities of classical solutions.
One of the approaches to this problem is called \emph{asymptotically safe gravity}~\cite{Reuter:2019byg}. It is crucially based on quantum field theory. A naïve way to study its phenomenology is via the renormalization group improvement~\cite{Bonanno:2001hi,Bonanno:2002zb,Rubano:2004jq,Koch:2013owa,Pawlowski:2018swz,Ishibashi:2021kmf,Chen:2022xjk}. Within this approach one integrates the beta function for the gravitational coupling only, neglecting the variations of other couplings. As a result of such an approximation one finds the effective Newton constant as a function of the energy scale $k$.
Then, this energy-dependent Newton constant is used in the classical black-hole solution in order to obtain the quantum corrected lapse function, describing a regular black hole~\cite{Bonanno:2000ep,Bonanno:2006eu,Falls:2010he,Falls:2012nd,Torres:2014gta,Torres:2014pea,Bonanno:2017zen,Adeifeoba:2018ydh,Held:2019xde}. Actually, the idea of energy dependence of the laws of physics at the level of the action is usual for quantum field theory.

Further, the gravitational coupling depends on some arbitrary renormalization group scale $k$, so that the relation connecting the energy and the radial coordinate must be defined in order to write down the resulting quantum corrected black hole metric. At this stage we are aware of two opportunities for the identification of the parameter $k$:
\begin{itemize}
\item with the modified proper distance, resulting in the so-called Bonanno-Reuter black hole metric \cite{Bonanno:2000ep}, and
\item with a power of the Kretschmann scalar~\cite{Pawlowski:2018swz,Adeifeoba:2018ydh}, leading to the Hayward metric~\cite{Held:2019xde}, which was initially suggested as a toy-model solution of the singularity problem within the collapse and evaporation scenario~\cite{Hayward:2005gi}.
\end{itemize}
A self-consistent deduction of the quantum corrected black-hole metric with the help of the renormalization group approach was suggested by Kazakov and Solodukhin in~\cite{Kazakov:1993ha}, and the quasinormal modes and grey-body factors were analyzed for this case in~\cite{Konoplya:2019xmn,Saleh:2016pke}. However, the Kazakov-Solodukhin solution is not free of singularity, but moves it to a finite distance from the center.

One of the basic characteristics of black holes, which depend only on the parameters of the background and not on the way of perturbations, are quasinormal modes~\cite{Konoplya:2011qq,Kokkotas:1999bd}. They dominate in the decay of perturbations at late times and, therefore, are observed by gravitational interferometers~\cite{ligo1,ligo2,ligo3}. Nevertheless, the uncertainty in the angular momentum and mass of the observed black holes leaves a wide window for modified theories of gravity~\cite{Konoplya:2016pmh}.

By now quasinormal modes of various regular black hole models have been extensively studied in a great number of papers~\cite{Bronnikov:2012ch,Lopez:2022uie,Lan:2022qbb,Rincon:2020cos,Jusufi:2020odz,Lan:2020fmn,Hendi:2020knv,MahdavianYekta:2019pol,Panotopoulos:2019qjk,Saleh:2018hba,Wu:2018xza,Li:2016oif,Fernando:2015fha,Toshmatov:2015wga,Li:2013fka,Lin:2013ofa,Flachi:2012nv}.
There is an extensive literature on the spectrum of the Hayward solution~\cite{Flachi:2012nv,Li:2013fka,Lin:2013ofa,Toshmatov:2015wga,Roy:2022rhv} and the quasinormal modes of the Bonanno-Reuter black hole have been studied in~\cite{Rincon:2020iwy,Panotopoulos:2020mii,Li:2013kkb} as well as its approximate, truncated version in~\cite{chinosII}. However in all the above studies a number of the crucial and, in our opinion, most interesting, properties of quasinormal spectra were missed for both cases.

First of all, the observed effect of quantum correction upon the fundamental mode reported in~\cite{Rincon:2020iwy} for the Bonanno-Reuter metric amounted to only a few percent. However, the analysis in~\cite{Rincon:2020iwy} does not include the whole parametric range and here we will show that the actual effect can be more than three times larger, exceeding $20$ percent for the extremal (Planckian) black hole configuration as compared to the corresponding Schwarzschild metric.

The most interesting behavior takes place for the overtones, which was also missed in the previous papers. Although it is believed that the major contribution to the signal is owing to the fundamental mode, recently it has been shown in~\cite{Giesler:2019uxc} (and later studied in~\cite{Oshita:2021iyn,Forteza:2021wfq}) that up to ten first overtones are necessary in order to model the ringdown of accurate numerical relativity simulations at all times and not only at the last stage. This finding indicated also that the actual quasinormal ringing starts earlier than expected. The main motivation of the renormalization group approach considered here is certainly related to quantum corrections, which are supposed to be small for astrophysical black holes. However, our finding may have broader interpretation, including large black holes and the aspects of the overtones' behavior might be useful. In addition, we will argue that some of the features found here must be generally appropriate to a quantum corrected black hole.

Here we will show that overtones for the above regular black holes have two peculiarities: First, there are purely imaginary modes in the spectra of both black holes. These non-oscillatory modes occur already at the second overtone and therefore may affect the ringdown at earlier times. Another interesting property is the high sensitivity of overtones to the near horizon asymptotic, which leads to enormous (up to hundreds of percent) deviation of the overtones' real parts from their Schwarzschild limit, while the fundamental mode deviates from its classical limit by (at most) a few percent. In other words, in the regime when the geometry of the quantum corrected black holes is very close to the Schwarzschild one everywhere except a small region near the event horizon, the overtones differ a lot from their Schwarzschild analogues. The above observation is apparently related to a similar phenomenon found in \cite{Jaramillo:2020tuu}, which was called\emph{ quasinormal modes' instability}: There it was observed that once there is a small perturbation of the initial linearized Einstein wave equations, say, for the Schwarzschild background, then while the fundamental mode and first few overtones are usually changed insignificantly, higher overtones are changed at a much greater degree. In other words, highly damped modes are sensitive to small deformations of the wave equation. In \cite{Jaramillo:2021tmt} it was shown that this quasinormal modes' instability might be visible in the gravitational signal in future.

Here we will give a systematic analysis of quasinormal ringing (including the overtones' behavior) of the above two regular black hole solutions for the test scalar, electromagnetic, and Dirac fields.

The analysis of gravitational perturbations for these vacuum solutions of the Einstein equation, could, in principle, be formally considered here as well.
However, the main supposition of the renormalization group approach is that the dominant correction to the background metric is due to the running gravitational coupling, what implies the existence of other correction terms which were neglected. Under this supposition there is no evidence that small perturbations of the black hole background spacetime will be still much smaller than the unknown terms we initially neglected in the background. In addition, we know that gravitational perturbations are usually qualitatively similar to those for test fields and frequently coincide with the latter in the high frequency (eikonal) regime.

Our paper is organized as follows: In sec.~\ref{Classical} we give the basic information on the quantum corrected black hole metrics under consideration. Sec.~\ref{WE} is devoted to the wave-like equation for the scalar, electromagnetic, and Dirac equations. In sec.~\ref{methods} we briefly review the Wentzel–Kramers–Brillouin (WKB), time-domain integration, and Frobenius methods used for calculations of quasinormal modes. In sec.~\ref{QNM} the quasinormal ringing, overtones and time-domain profiles are analyzed for both metrics and the analytical formula for quasinormal frequencies is deduced for the eikonal (high multipole number) regime. Finally in sec.~\ref{Conclusions} we summarize the obtained results and discuss some open problems.

\section{Renormalization group-improvement for black hole solutions}\label{Classical}
The improved Schwarzschild metric starts from the average (ghost-free) Einstein-Hilbert action in the four dimensional spacetime \cite{Bonanno:2000ep}:
\begin{equation}\nonumber
\Gamma_k[g_{\mu\nu}] = \frac{1}{16 \pi G(k)} \int \mathrm{d}^4 x \sqrt{\det(g_{\mu\nu})} \Bigl( - R(g_{\mu\nu}) + 2 \bar{\lambda}(k)\Bigl).
\end{equation}
The scale-dependent couplings are described by the Wetterich equation
\begin{equation}
\partial_{k}\Gamma_{k} = \frac{1}{2} \text{Tr}\left[\left(\Gamma^{(2)}_{k} + \mathcal{R}_{k}\right)^{-1}\cdot\partial_{k}\mathcal{R}_{k}\right],
\end{equation}
where $\Gamma_{k}^{(2)}$ stands for the Hessian of $\Gamma_k$ with respect to $g_{\mu \nu}$, and $\mathcal{R}_k$ is a cutoff function which implements the infrared cutoff, as was pointed out in Ref.~\cite{Bonanno:2000ep}.

Following \cite{Bonanno:2000ep}, the evolution of the dimensionless gravitational coupling is described by the following equation
\begin{equation} \label{betag}
k\frac{\mathrm{d}}{\mathrm{d}k} k^2 G(k) = \left[ 2 + \frac{B_1 k^2 G(k)}{1- B_2 k^2 G(k)} \right] \: k^2 G(k),
\end{equation}
where $B_1$ and $B_2$ are some factors. Here we assume that the dimensionless running cosmological constant
$$\lambda = \bar{\lambda}/k^2 \ll 1 $$
for all scales of interest, so that we may approximate $\lambda \approx 0$ in the arguments of $B_{1}(\lambda)$ and $B_{2}(\lambda)$.

Integrating the equation~\eqref{betag}, one can find the Newton's coupling
\begin{align} \label{basic}
G(k) &= \frac{G_0}{1 + \tilde{\omega} G_0 k^2},
\end{align}
where $\tilde{\omega}$ is a constant.

From the above relation we can see that the effects of quantum corrections are essential at high energies and at the low ones the classical solution is reproduced.
The next step in the procedure is the identification of $k$ with some physical parameter. At this stage there are various ways to identify the parameter $k$ in the literature, of which we will consider the following two:
\begin{enumerate}
\item $k(r)$ is a modified proper distance, which leads to the Bonanno-Reuter metric~\cite{Bonanno:2000ep};
\item $k(r) = \alpha K^{-1/4}$, where $K$ is the Kretschmann scalar and $\alpha = 48^{-1/4}$~\cite{Held:2019xde,Pawlowski:2018swz,Adeifeoba:2018ydh}), coinciding, up to the redefinition of constants, with the Hayward metric~\cite{Hayward:2005gi}.
\end{enumerate}

The metric of a spherically-symmetric black hole has the following general form
\begin{equation}
\mathrm{d}s^2 = -f(r)\mathrm{d} t^2 + f(r)^{-1} \mathrm{d} r^2 + r^2 \: \mathrm{d}\Omega^2,
\end{equation}
with the only difference that now the Newton's ``constant'' depends on $r$.

The metric function for the Bonanno-Reuter identification (supposing that in the classical limit $G_0=1$) has the form
\begin{equation}
f(r) = 1-\frac{2 M r^2}{r^3 + \frac{118}{15 \pi} \left(r + \frac{9}{2} M \right)},
\end{equation}
where $M$ is the black hole mass, measured in units of the Planck mass,
$$m_P\equiv\sqrt{\hbar c/G_0}\approx2.176\times10^{-8}\mathrm{kg}\approx1.6\times10^{-35}\mathrm{m}.$$
The event horizon exists when $M \geq M_{\text{crit}}$,
where
$$\frac{M_{\text{crit}}}{m_P}=\frac{23 \sqrt{5}+ 5 \sqrt{221}}{720} \sqrt{\frac{59}{3\pi}\left(31 + \sqrt{1105}\right)}\approx 3.503.$$

The Hayward metric function has the form:
\begin{equation}
f(r) = 1-\frac{2 r^2/M^2}{r^{3}/M^{3}+ \gamma},
\end{equation}
where the event horizon exists whenever $\gamma \leq 32/27$.

\section{The wave-like equations}\label{WE}
The general covariant equations for the scalar field $\Phi$, the electromagnetic field $A_\mu$, and the Dirac field $\Upsilon$ \cite{Brill:1957fx} are respectively written as:
\begin{subequations}\label{coveqs}
\begin{eqnarray}\label{KGg}
\frac{1}{\sqrt{-g}}\partial_\mu \left(\sqrt{-g}g^{\mu \nu}\partial_\nu\Phi\right)&=&0,
\\\label{EmagEq}
\frac{1}{\sqrt{-g}}\partial_{\mu} \left(F_{\rho\sigma}g^{\rho \nu}g^{\sigma \mu}\sqrt{-g}\right)&=&0\,,
\\\label{covdirac}
\gamma^{\alpha} \left( \frac{\partial}{\partial x^{\alpha}} - \Gamma_{\alpha} \right) \Upsilon&=&0,
\end{eqnarray}
\end{subequations}
where $F_{\mu\nu}=\partial_\mu A_\nu-\partial_\nu A_\mu$ is the electromagnetic tensor, $\gamma^{\alpha}$ are noncommutative gamma matrices and $\Gamma_{\alpha}$ are spin connections in the tetrad formalism.
After separation of the variables equations (\ref{coveqs}) take the Schrödinger-like form (see, for instance, \cite{Konoplya:2011qq,Kokkotas:1999bd} and references therein)
\begin{equation}\label{wave-equation}
\dfrac{d^2 \Psi}{dr_*^2}+(\omega^2-V(r))\Psi=0,
\end{equation}
where the ``tortoise coordinate'' $r_*$ is defined by the following relation
\begin{equation}
dr_*\equiv\frac{dr}{f(r)}.
\end{equation}

The effective potentials for the scalar ($s=0$) and electromagnetic ($s=1$) fields are
\begin{equation}\label{potentialScalar}
V(r)=f(r) \frac{\ell(\ell+1)}{r^2}+\left(1-s\right)\cdot\frac{f(r)}{r}\frac{d f(r)}{dr},
\end{equation}
where $\ell=s, s+1, s+2, \ldots$ are the multipole numbers.
For the Dirac field ($s=1/2$) we have two iso-spectral potentials
\begin{equation}
V_{\pm}(r) = W^2\pm\frac{dW}{dr_*}, \quad W\equiv \left(\ell+\frac{1}{2}\right)\frac{\sqrt{f(r)}}{r}.
\end{equation}
The iso-spectral wave functions can be transformed one into another by the Darboux transformation
\begin{equation}\label{psi}
\Psi_{+}=q \left(W+\dfrac{d}{dr_*}\right) \Psi_{-}, \quad q=const.
\end{equation}
Therefore, we will calculate quasinormal modes for only one of the effective potentials, $V_{+}(r)$, because the WKB method works better for it.

\section{Methods used for calculations of quasinormal modes}\label{methods}
Quasinormal modes $\omega_{n}$ are proper oscillation frequencies corresponding to the solutions of the master wave equation (\ref{wave-equation}) when the purely outgoing waves at both infinities are imposed:
\begin{equation}
\Psi \propto e^{-\imo \omega t \pm \imo \omega r_*}, \quad r_* \to \pm \infty.
\end{equation}
Thus, there are no waves coming from either infinity or the event horizon. Here we will review the three methods (time-domain integration, WKB, and Frobenius) used for calculations of quasinormal frequencies.

\subsection{Time-domain integration}
In order to find quasinormal modes and, foremost, analyze possible echo-like phenomena we will use the time-domain integration method. We will integrate the wavelike equation (\ref{wave-equation}) in terms of the light-cone variables $u=t-r_*$ and $v=t+r_*$ via applying the discretization scheme of Gundlach-Price-Pullin \cite{Gundlach:1993tp},
\begin{eqnarray}\label{Discretization}
\Psi\left(N\right)&=&\Psi\left(W\right)+\Psi\left(E\right)-\Psi\left(S\right) \PRDstyle{\\\nonumber&&}
-\Delta^2V\left(S\right)\frac{\Psi\left(W\right)+\Psi\left(E\right)}{4}+{\cal O}\left(\Delta^4\right)\,,
\end{eqnarray}
where the following notation for the points was used:
$N\equiv\left(u+\Delta,v+\Delta\right)$, $W\equiv\left(u+\Delta,v\right)$, $E\equiv\left(u,v+\Delta\right)$, and $S\equiv\left(u,v\right)$. The Gaussian initial data are imposed on the two null surfaces, $u=u_0$ and $v=v_0$. The dominant quasinormal frequencies can be extracted from the time-domain profiles with the help of the Prony method see, e.g.,~\cite{Konoplya:2011qq}.

\subsection{WKB method}
In the frequency domain we will use the semi-analytic WKB method applied by Will and Schutz \cite{Schutz:1985km} for finding quasinormal modes. The Will-Schutz formula was extended to higher orders in \cite{Iyer:1986np,Konoplya:2003ii,Matyjasek:2017psv} and made even more accurate when using the Padé approximants in \cite{Matyjasek:2017psv,Hatsuda:2019eoj}.
The higher-order WKB formula has the form \cite{Konoplya:2019hlu},
\begin{eqnarray}
\omega^2&=&V_0+A_2(\K^2)+A_4(\K^2)+A_6(\K^2)+\ldots \\\nonumber
&-& \imo \K\sqrt{-2V_2}\left(1+A_3(\K^2)+A_5(\K^2)+A_7(\K^2)+\ldots\right),
\end{eqnarray}
where $\K=n+1/2$ is half-integer. The corrections $A_k(\K^2)$ of the order $k$ to the eikonal formula are polynomials of $\K^2$ with rational coefficients and depend on the values of higher derivatives of the potential $V(r)$ in its maximum. In order to increase accuracy of the WKB formula, we will follow the procedure of Matyjasek and Opala \cite{Matyjasek:2017psv} and use the Padé approximants. Here we will use the sixth order WKB method with $\tilde{m} =6$, where $\tilde{m}$ is defined in \cite{Matyjasek:2017psv,Konoplya:2019hlu}, because this choice provides the best accuracy in the Schwarzschild limit.

\subsection{Frobenius method}
In order to find the accurate values of QNMs we employ the method proposed by Leaver~\cite{Leaver:1985ax}. For the black holes of Bonanno-Reuter and Hayward the wave-like equation~(\ref{wave-equation}) has regular singularities at the event horizon $r=r_+$, the inner horizon $r=r_-$, and the irregular singularity at spatial infinity $r=\infty$. We introduce the new function,
\begin{equation}\label{reg}
\Psi(r)=e^{\imo\omega r}(r-r_-)^{\lambda}\left(\frac{r-r_+}{r-r_-}\right)^{-\imo\omega/f'(r_+)}y(r),
\end{equation}
where $\lambda$ is defined in such a way that $y(r)$ is regular at $r=\infty$ once $\Psi(r)$ corresponds to the purely outgoing wave at spatial infinity. We notice that $\Psi(r)$ behaves as the purely ingoing wave at the event horizon if $y(r)$ is regular at $r=r_+$. Therefore, we represent $y(r)$ in terms of the following Frobenius series:
\begin{equation}\label{Frobenius}
y(r)=\sum_{k=0}^{\infty}a_k\left(\frac{r-r_+}{r-r_-}\right)^k,
\end{equation}
and find that the coefficients $a_k$ satisfy the 10-term recurrence relation, which can be reduced to the three-term recurrence relation via Gaussian elimination (see \cite{Konoplya:2011qq} for a detailed description of the procedure). Finally, using the recurrence relation coefficients, we find the equation with the infinite continued fraction with respect to $\omega$, which is satisfied iff the series (\ref{Frobenius}) converges at $r=\infty$, i.e., when $\Psi(r)$ obeys the quasinormal boundary conditions. In order to calculate the infinite continued fraction we use the Nollert improvement~\cite{Nollert:1993zz}, which was generalized in~\cite{Zhidenko:2006rs} for an arbitrary number of terms in the recurrence relation.

\section{Quasinormal frequencies and time-domain profiles}\label{QNM}

Here we will use all the three above methods in order to find quasinormal modes. The fundamental mode can be found with sufficient accuracy by the WKB method and time-domain integration, while the overtones, specially those for which $n > \ell$, have been calculated with the Frobenius method.

\begin{figure*}
\resizebox{\linewidth}{!}{\includegraphics{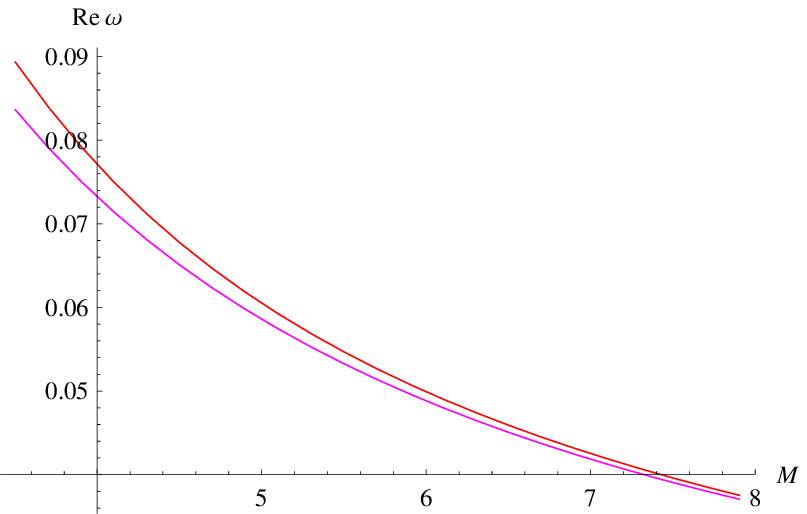}\includegraphics{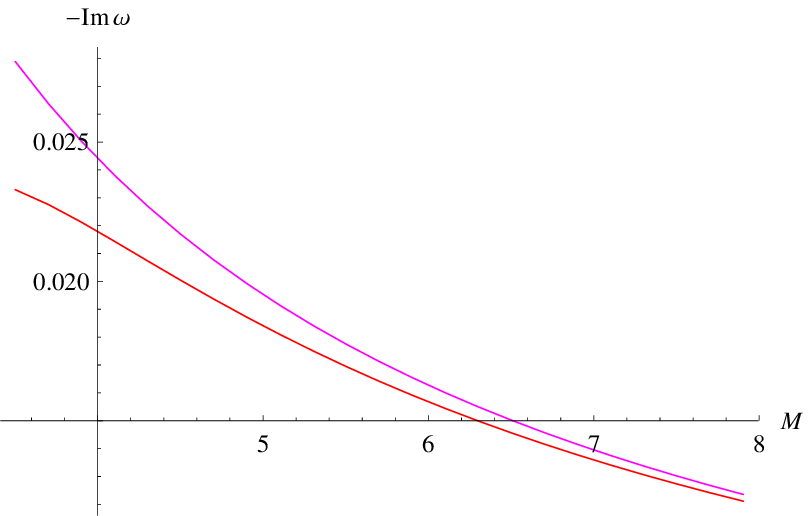}}
\caption{The fundamental ($n=0$) quasinormal modes of the Bonanno-Reuter black hole model (red) versus Schwarzschild modes (magenta) with the same mass for $\ell=1$ scalar perturbations computed by the WKB method. The Schwarzschild mode has lower $\re{\omega}$ and higher decay rate. }\label{fig1}
\end{figure*}

\begin{figure*}
\resizebox{\linewidth}{!}{\includegraphics{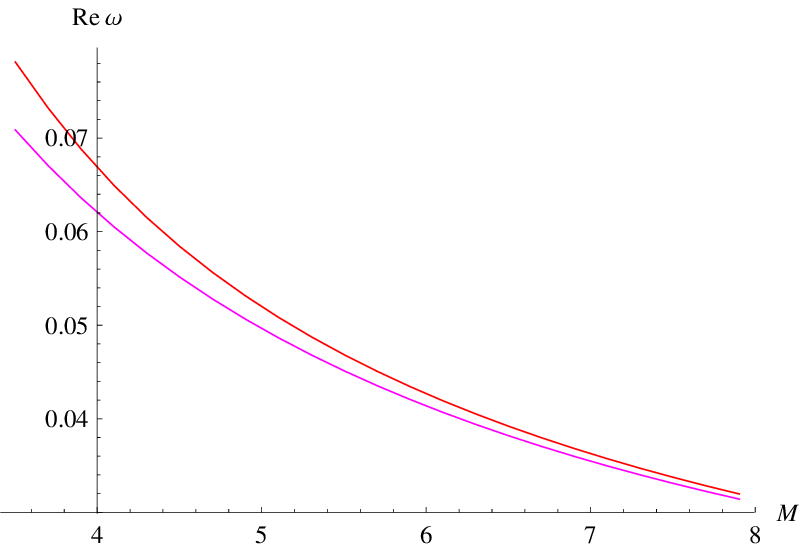}\includegraphics{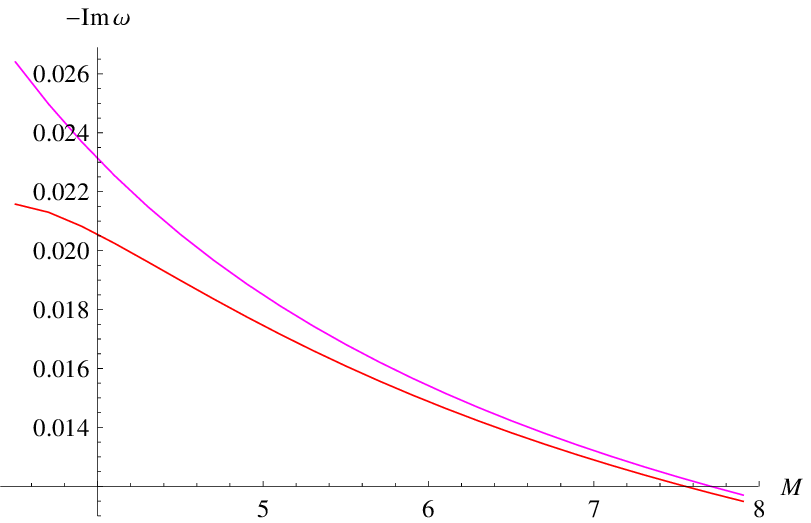}}
\caption{The fundamental ($n=0$) quasinormal modes of the Bonanno-Reuter black hole model (red) versus Schwarzschild modes (magenta) with the same mass for $\ell=1$ electromagnetic perturbations computed by the WKB method. The Schwarzschild mode has lower $\re{\omega}$ and higher decay rate.}\label{fig2}
\end{figure*}

\begin{figure*}
\resizebox{\linewidth}{!}{\includegraphics{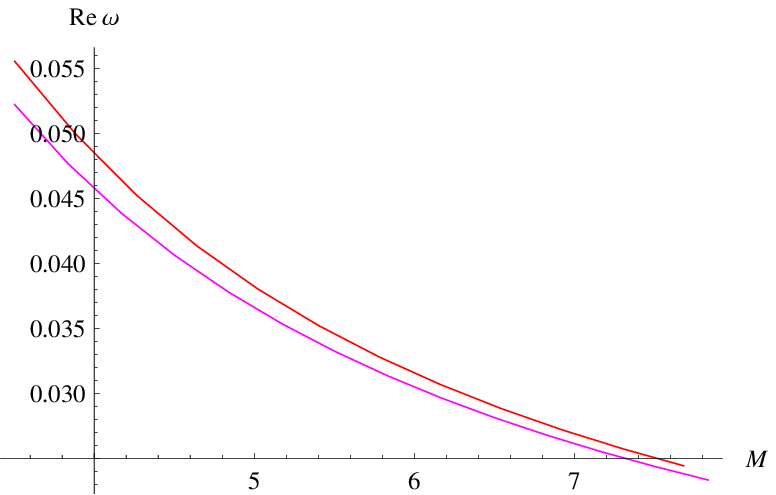}\includegraphics{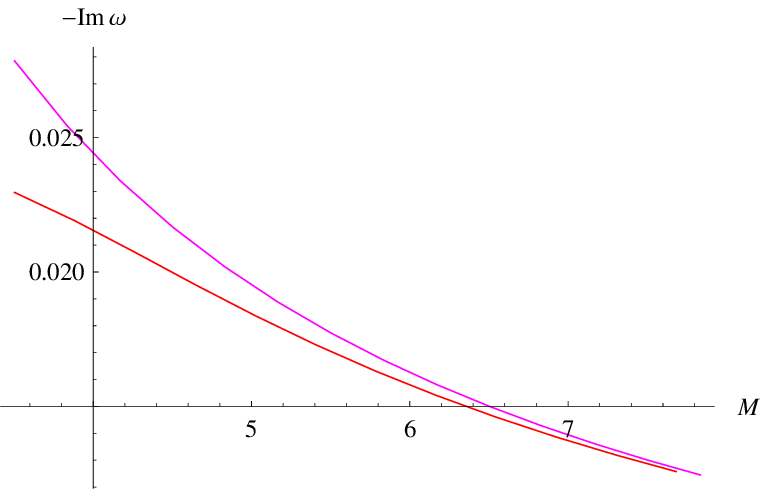}}
\caption{The fundamental ($n=0$) quasinormal modes of the Bonanno-Reuter black hole model (red) versus Schwarzschild modes (magenta) with the same mass for $\ell=1/2$ Dirac perturbations computed by the WKB method. The Schwarzschild mode has lower $\re{\omega}$ and higher decay rate.}\label{fig3}
\end{figure*}

From figs.~\ref{fig1}, \ref{fig2}, and~\ref{fig3} one can see that the real oscillation frequency of the fundamental mode is larger for the Bonanno-Reuter black hole than for the Schwarzschild one with the same mass for scalar, electromagnetic, and Dirac fields.
On the opposite, the damping rate is smaller than that of the Schwarzschild black hole. In the regime of large mass the quasinormal frequencies of the Bonanno-Reuter spacetime approach their Schwarzschild limit. In \cite{Rincon:2020iwy} it was claimed that the effect of the Bonanno-Reuter correction results in only a few percent difference from the Schwarzschild spectrum. However, there the minimal value of mass was taken $M=5$, which is different from the extremal value $M \approx 3.503$ (in Planck units). Here we extended the calculations to the extremal limit and found that, even for the fundamental mode, the effect is much larger than previously reported and exceeds $20$ percent.

\begin{figure*}
\resizebox{\linewidth}{!}{\includegraphics{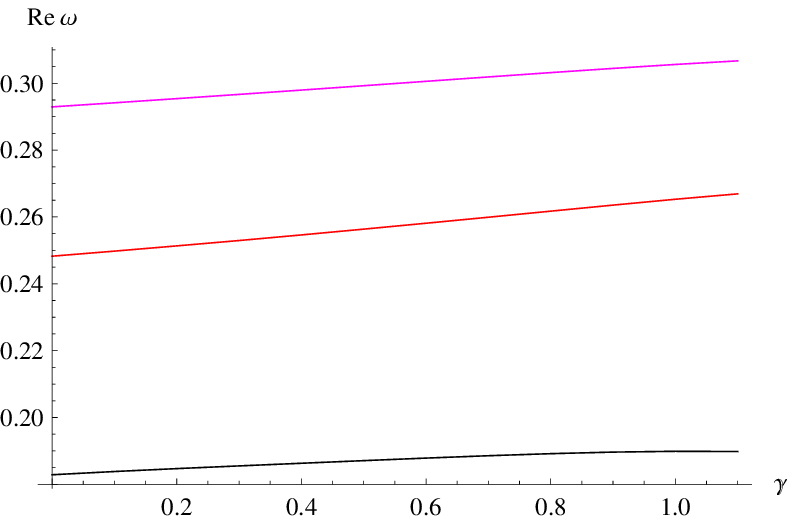}\includegraphics{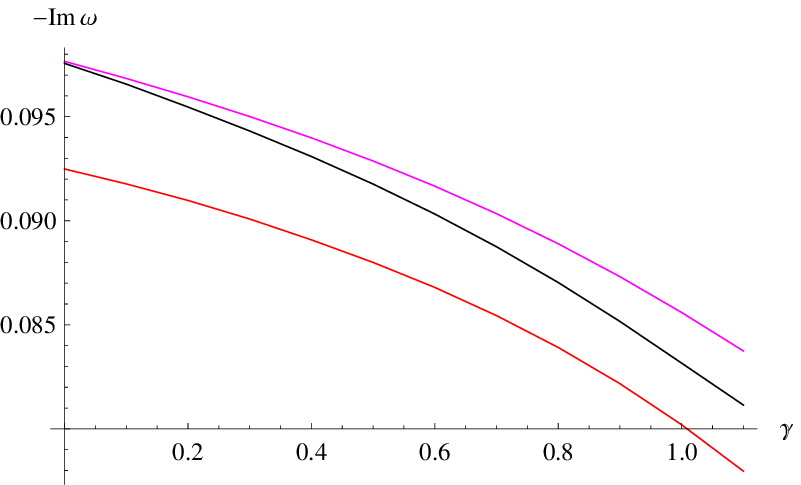}}
\caption{The fundamental ($n=0$) quasinormal modes of the Hayward black hole model for $\ell=1$ scalar (magenta, upper) and electromagnetic (red, lower imaginary part) perturbations, and $\ell=1/2$ Dirac (black, lower real part) perturbations computed by the WKB method.}\label{fig4}
\end{figure*}

From fig.~\ref{fig4} we can see that the Hayward black hole is characterized by qualitatively the same behavior of the fundamental mode: renormalization group correction leads to higher real oscillation frequency and slower decay. Exception from such monotonic behavior occurs only for the $\ell=0$ scalar field perturbation at $\gamma \approx 20/27$ for the fundamental mode (see table \ref{table4}). Overall, we can conclude that the quality factor of the corrected black hole is higher in both cases, that is, a black hole in asymptotically safe gravity is a much better oscillator than its classical limit.

Comparison of the WKB data with the time-domain integration and Frobenius method (see tables~\ref{table1}, \ref{table2}, and~\ref{table4}) shows excellent concordance in the whole range of parameters. The Frobenius method is accurate, because it is based on the convergent procedure. This means that it is the criterium of accuracy for the other two methods and consequently both the time-domain integration and the WKB formula provide reasonable accuracy at least for the fundamental mode. For the near extreme Hayward black hole the Frobenius method converges very slowly, so that the extremal limit in table \ref{table4} is calculated only by the time-domain integration and WKB methods.

Using the first order WKB formula we can obtain quasinormal modes in the high multipole number $\ell\to\infty$ regime in analytic form. For this we will use the expression for the position of the peak of the effective potential, which for the Bonanno-Reuter metric is located at:
\begin{eqnarray}
r_{max} &=& 3 M-\frac{1652}{135 \pi M} -\frac{5071817}{54675 \pi ^2 M^3} \PRDstyle{\\\nonumber&&}
-\frac{28261793432}{22143375 \pi ^3 M^5} - \frac{64480973421931}{2989355625 \pi ^4 M^7}.
\end{eqnarray}
Then, the WKB formula yields
\begin{eqnarray}
\im{\omega} &=& \frac{\left(n+\frac{1}{2} \right)}{3 \sqrt{3} M} \left(1-\frac{1298}{405 \pi M^2} + \mathcal{O}\left(\frac{1}{M^{4}}\right) \right), \\
\re{\omega} &=& \frac{\left(\ell+\frac{1}{2}\right)}{3 \sqrt{3} M} \left(1+\frac{59}{27 \pi M^2} + \mathcal{O}\left(\frac{1}{M^{4}}\right) \right).
\end{eqnarray}
The same procedure for the Hayward metric leads to the following expressions for the maximum of the potential,
\begin{equation}
r_{max} = 3 M -\frac{2 \gamma M}{9} -\frac{\gamma ^2 M}{27} -\frac{70 \gamma ^3 M}{6561} -\frac{665 \gamma ^4 M}{177147},
\end{equation}
and eikonal quasinormal modes,
\begin{eqnarray}
\im{\omega} &=& \frac{\left(n+\frac{1}{2} \right)}{3 \sqrt{3} M} \left(1-\frac{2}{27} \gamma + \mathcal{O}(\gamma^2) \right), \\
\re{\omega} &=& \frac{\left(\ell+\frac{1}{2} \right)}{3 \sqrt{3} M} \left(1-\frac{2}{54} \gamma + \mathcal{O}(\gamma^2) \right).
\end{eqnarray}
Notice, that according to \cite{Cardoso:2008bp} there is a correspondence between eikonal quasinormal modes of test fields and characteristics of null geodesics: The real and imaginary parts of $\omega$ are multiples of the frequency and instability timescale of the circular null geodesics respectively. This correspondence, however, is not guaranteed for gravitational and other non-minimally coupled fields \cite{Konoplya:2017wot,Konoplya:2019hml}.

\begin{figure*}
\resizebox{\linewidth}{!}{\includegraphics{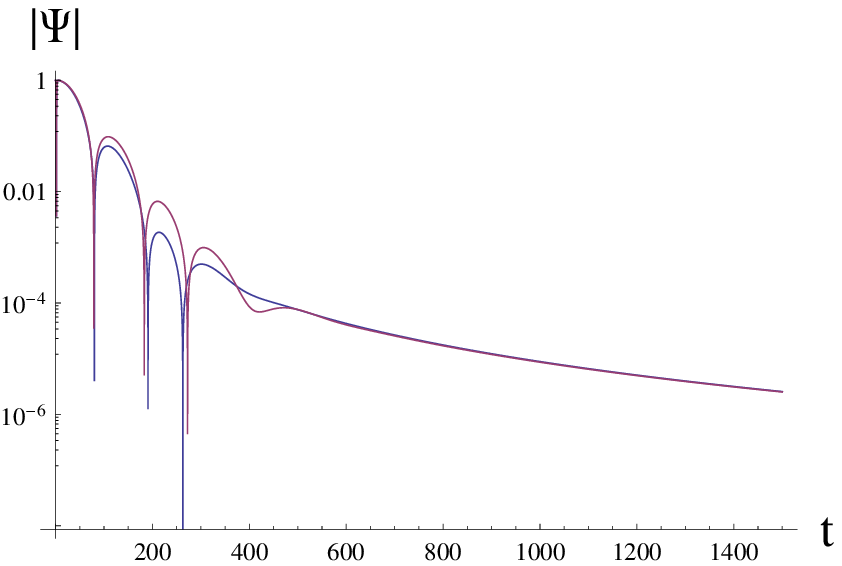}\includegraphics{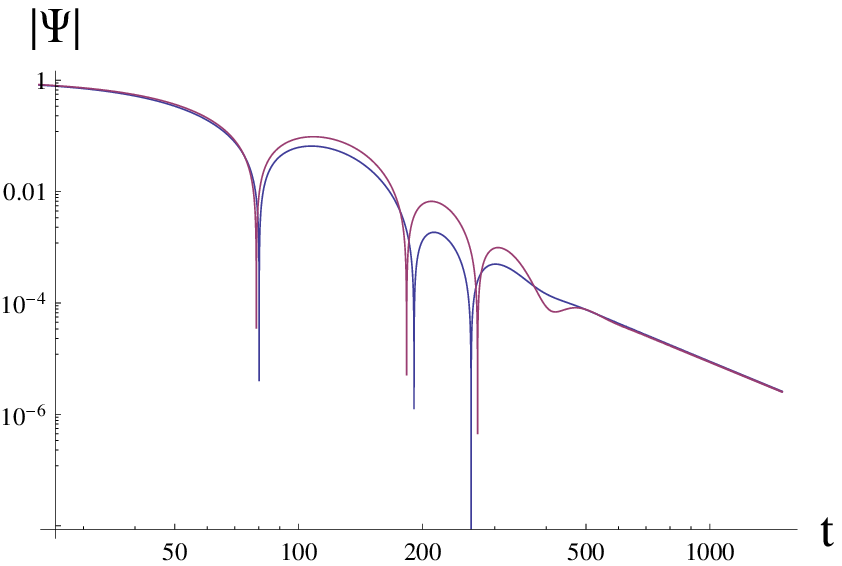}}
\caption{Time-domain profile of $\ell=0$ scalar perturbations of the Bonanno-Reuter (magenta, upper) and Schwarzschild (blue, lower) black holes, $M=3.56$: Left panel -- semi-logarithmic plot, right panel -- logarithmic plot. }\label{TDfig1}
\end{figure*}

\begin{figure*}
\resizebox{\linewidth}{!}{\includegraphics{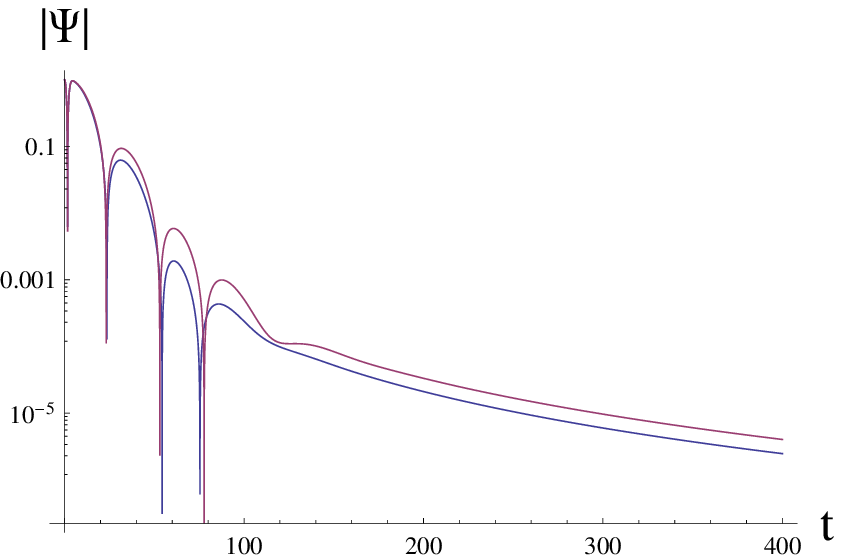}\includegraphics{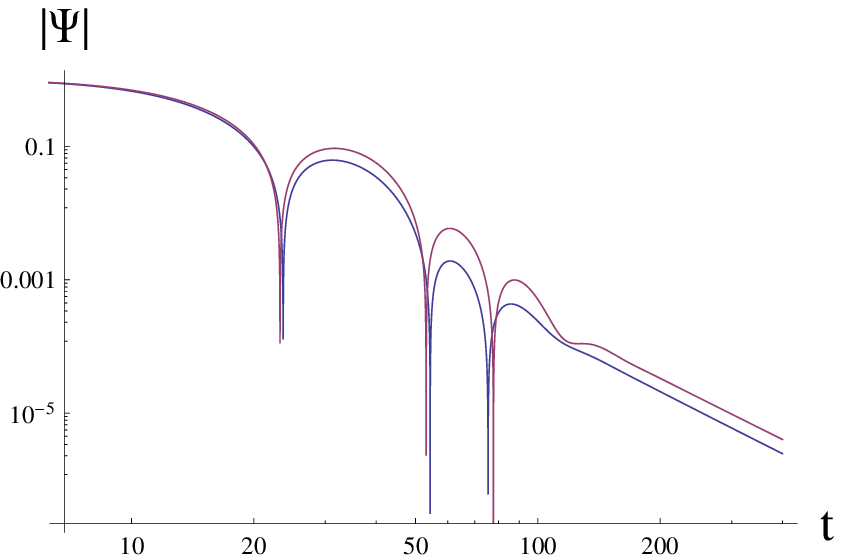}}
\caption{Time-domain profile of $\ell=0$ scalar perturbations of the Hayward (magenta, upper) and Schwarzschild (blue, lower) black holes, $\gamma=32/27$: Left panel -- semi-logarithmic plot, right panel -- logarithmic plot. }\label{TDfig2}
\end{figure*}

From figs.~\ref{TDfig1}~and~\ref{TDfig2} we can see time-domain profiles of evolution of perturbations for the Bonanno-Reuter and Hayward metrics in comparison with that for the Schwarzschild solution. While the initial period of quasinormal ringing is characterized by relatively close fundamental quasinormal modes for both Schwarzschild and the corresponding corrected Schwarzschild-like metric, with time the initial phase is changed and the profiles do not merge. At asymptotically late times quasinormal modes are suppressed by the power-law tails which are the same as for the Schwarzschild case. Thus, for example, for the scalar perturbations shown on the above figures we have:
\begin{equation}
\Psi \propto t^{-(2 \ell +3)}, \quad t \to \infty.
\end{equation}

The tails in fig.~\ref{TDfig1} almost merge, because the geometries of the Schwarzschild and Bonanno-Reuter metrics are almost indistinguishable in the regime of large mass $M$ chosen there. On the contrary, tails are parallel, but do not merge in fig.~\ref{TDfig2}, because the parameter $\gamma$ there is fixed to its extreme value and the deviation from the Schwarzschild metric is large.

The most striking difference occurs not at the fundamental mode, but at the overtones. First of all, for the Bonanno-Reuter black hole we have observed that the overtones differ from the Schwarzschild spectrum more than the fundamental mode: The real part of the higher overtones is much smaller than the corresponding mode of the Schwarzschild black hole. Second, we have found a purely imaginary mode for both test scalar $\ell=0$, $n=7$ (see table~\ref{table2}) and electromagnetic $\ell=1$, $n=6$ (table~\ref{table2e}) fields when $M=4$. At larger mass, this non-oscillatory mode moves to higher overtones, being already $n=11$ mode for $M=8$. Notice, that the next overtone $n=12$ is also purely imaginary at least within the numerical precision of 6 decimal places. Thus, in general there are more than one purely imaginary mode. When the mass is further increased, the purely imaginary modes move to even higher overtone number and simply disappear from the spectrum in the Schwarzschild limit. For the larger multipole we observe qualitatively the same effect: the real part rapidly decreases with the overtone number. However, since the real part grows with the multipole number, the purely imaginary mode usually occurs at larger overtones.

\begin{figure*}
\resizebox{\linewidth}{!}{\includegraphics{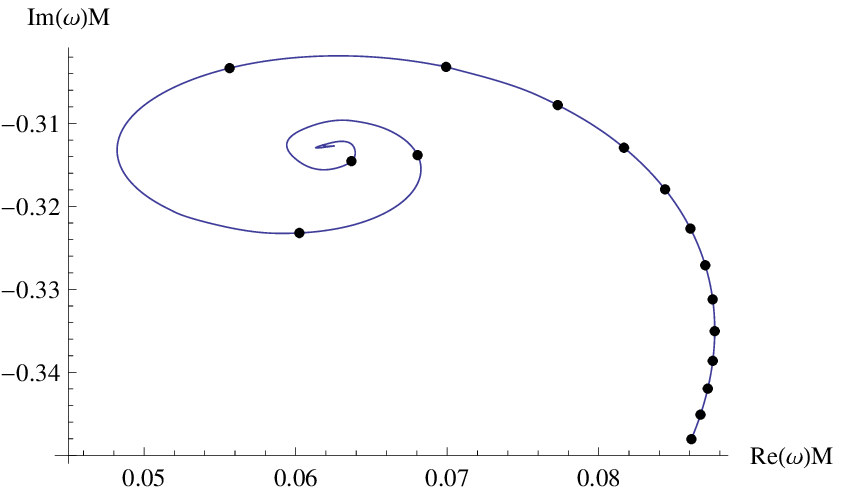}\includegraphics{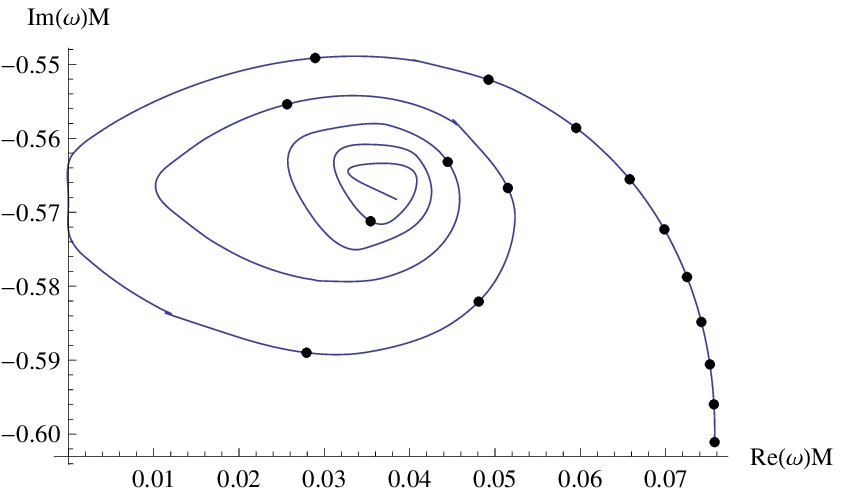}}
\caption{First ($n=1$) and second ($n=2$) overtones of $\ell=0$ scalar perturbations of the Hayward black holes parametrized by $\gamma$: The black dots correspond to $27\gamma=0,2,4,\ldots,30$ (from the bottom right corner, corresponding to the Schwarzschild mode, to the center of the spiral-like curve). For the parameters $0.702\lesssim\gamma\lesssim0.705$, the real part of the second overtone (right panel) becomes close to zero. The real part of the first overtone (left panel) is nonzero for all values of $\gamma$.}\label{overtones}
\end{figure*}

The scalar and electromagnetic fields in the Hayward background show a similar behavior: The real part of the overtones is suppressed as compared to the Schwarzschild black hole, and the non-oscillatory purely (or almost purely) damped modes appear in the spectra of the test fields. For $\gamma=1$ the purely imaginary modes are $\ell=0$, $n=4$ for the scalar field (table~\ref{table5}) and $\ell=1$, $n=6$ for the electromagnetic field (table~\ref{table5e}). The overtones' behavior for the Hayward black hole is illustrated in fig.~\ref{overtones}. One can see that, unlike the dominant mode (see table~\ref{table4}), the first overtone as a function of $\gamma$ goes along a spiral-like curve, though within a bounded region, such that only the real part changes significantly (within about $50\%$ of the Schwarzschild value). For the second overtone the spiraling occurs faster as $\gamma$ grows and the real part vanishes for some values of $\gamma$. Thus, one can observe the non-oscillating modes already at the second overtone.

\begin{table*}
\PRDstyle{\begin{tabular}{c@{\hspace{1em}}|@{\hspace{1em}}c@{\hspace{2em}}c@{\hspace{2em}}c@{\hspace{1em}}|@{\hspace{1em}}c}}
\JCAPstyle{\begin{tabular}{c|ccc|c}}
  \hline
  $M$ & $\omega$ (WKB) & $\omega$ (time-domain) & $\omega$ (accurate) & $\omega$ (Schwarzschild)\\
  \hline
  $3.503$ & $0.032153-0.025011\imo$ & $0.032413-0.025440\imo$ & $0.032094-0.025008\imo$ & $0.031532-0.029945\imo$ \\
  $4.503$ & $0.025774-0.021206\imo$ & $0.025602-0.021954\imo$ & $0.025759-0.021170\imo$ & $0.024529-0.023295\imo$ \\
  $5.503$ & $0.020744-0.018051\imo$ & $0.019615-0.018257\imo$ & $0.020820-0.018056\imo$ & $0.020072-0.019062\imo$ \\
  $6.503$ & $0.017421-0.015540\imo$ & $0.016514-0.015338\imo$ & $0.017450-0.015567\imo$ & $0.016985-0.016130\imo$ \\
  $7.503$ & $0.015030-0.013587\imo$ & $0.014592-0.012703\imo$ & $0.015027-0.013630\imo$ & $0.014721-0.013981\imo$ \\
  $8.503$ & $0.013216-0.012054\imo$ & $0.012958-0.011343\imo$ & $0.013201-0.012103\imo$ & $0.012990-0.012336\imo$ \\
  $9.503$ & $0.011793-0.010827\imo$ & $0.011683-0.010191\imo$ & $0.011775-0.010874\imo$ & $0.011623-0.011038\imo$ \\
  \hline
\end{tabular}
\caption{Fundamental quasinormal modes for the scalar $\ell=n=0$ perturbations of the Bonanno-Reuter black hole.}\label{table1}
\end{table*}

\begin{table*}
\begin{tabular}{c@{\hspace{1em}}|@{\hspace{1em}}c@{\hspace{1em}}c@{\hspace{1em}}|@{\hspace{1em}}c@{\hspace{1em}}c}
\hline
&\multicolumn{2}{c|@{\hspace{1em}}}{$\ell=0$}&\multicolumn{2}{c}{$\ell=1$}\\
\hline
  $n$ & $\omega$ (Bonanno-Reuter) & $\omega$ (Schwarzschild) & $\omega$ (Bonanno-Reuter) & $\omega$ (Schwarzschild) \\
\hline
  $0$ & $0.029000-0.022799\imo$ & $0.027614-0.026224\imo$ & $0.077094-0.021791\imo$ & $0.073234-0.024415\imo$ \\
  $1$ & $0.014735-0.075427\imo$ & $0.021529-0.087013\imo$ & $0.070283-0.067061\imo$ & $0.066112-0.076564\imo$ \\
  $2$ & $0.013708-0.140628\imo$ & $0.018936-0.150270\imo$ & $0.058062-0.116660\imo$ & $0.057385-0.135033\imo$ \\
  $3$ & $0.006542-0.207654\imo$ & $0.017603-0.213419\imo$ & $0.043693-0.174315\imo$ & $0.050815-0.197075\imo$ \\
  $4$ & $0.006835-0.263913\imo$ & $0.016769-0.276408\imo$ & $0.031041-0.236051\imo$ & $0.046277-0.260191\imo$ \\
  $5$ & $0.004802-0.331462\imo$ & $0.016185-0.339285\imo$ & $0.025512-0.299442\imo$ & $0.043019-0.323530\imo$ \\
  $6$ & $0.001296-0.388573\imo$ & $0.015749-0.402085\imo$ & $0.019751-0.366931\imo$ & $0.040557-0.386860\imo$ \\
  $7$ & $0.002675-0.455550\imo$ & $0.015405-0.464832\imo$ & $0.008952-0.427228\imo$ & $0.038614-0.450123\imo$ \\
  $8$ & $0.00-0.51\imo$         & $0.015127-0.527538\imo$ & $0.012764-0.492353\imo$ & $0.037032-0.513309\imo$ \\
  $9$ & $0.001631-0.579538\imo$ & $0.014894-0.590215\imo$ & $0.008551-0.560693\imo$ & $0.035709-0.576425\imo$ \\
\hline
\end{tabular}
\caption{Dominant quasinormal modes for the scalar perturbations of the Bonanno-Reuter black hole ($M=4$) and the corresponding modes for the Schwarzschild black hole.}\label{table2}
\end{table*}

\begin{table}
\begin{tabular}{c@{\hspace{1em}}|@{\hspace{1em}}c@{\hspace{2em}}c}
\hline
  $n$ & $\omega$ (Bonanno-Reuter) & $\omega$ (Schwarzschild) \\
\hline
  $0$ & $0.066877-0.020549\imo$ & $0.062066-0.023122\imo$ \\
  $1$ & $0.059764-0.063499\imo$ & $0.053629-0.073417\imo$ \\
  $2$ & $0.047392-0.110619\imo$ & $0.043693-0.131297\imo$ \\
  $3$ & $0.030448-0.166115\imo$ & $0.036544-0.192977\imo$ \\
  $4$ & $0.019951-0.224263\imo$ & $0.031639-0.255638\imo$ \\
  $5$ & $0.012209-0.290487\imo$ & $0.028063-0.318481\imo$ \\
  $6$ & $0.000983-0.354117\imo$ & $0.025304-0.381317\imo$ \\
  $7$ & $0.000-0.420\imo$       & $0.023081-0.444100\imo$ \\
\hline
\end{tabular}
\caption{Dominant quasinormal modes for the electromagnetic perturbations ($\ell=1$) of the Bonanno-Reuter black hole ($M=4$) and the corresponding modes for the Schwarzschild black hole.}\label{table2e}
\end{table}

\begin{table}
\begin{tabular}{c@{\hspace{1em}}|@{\hspace{1em}}c@{\hspace{2em}}c}
\hline
  $n$ & $\omega$ (Bonanno-Reuter) & $\omega$ (Schwarzschild) \\
\hline
  $0$ & $0.014060-0.012828\imo$ & $0.013807-0.013112\imo$ \\
  $1$ & $0.011030-0.042326\imo$ & $0.010765-0.043507\imo$ \\
  $2$ & $0.009447-0.073078\imo$ & $0.009468-0.075135\imo$ \\
  $3$ & $0.008318-0.103855\imo$ & $0.008801-0.106710\imo$ \\
  $4$ & $0.007282-0.134637\imo$ & $0.008384-0.138204\imo$ \\
  $5$ & $0.006213-0.165471\imo$ & $0.008093-0.169642\imo$ \\
  $6$ & $0.005055-0.196409\imo$ & $0.007874-0.201043\imo$ \\
  $7$ & $0.003780-0.227529\imo$ & $0.007703-0.232416\imo$ \\
  $8$ & $0.002403-0.258950\imo$ & $0.007563-0.263769\imo$ \\
  $9$ & $0.001041-0.290822\imo$ & $0.007447-0.295107\imo$ \\
 $10$ & $0.000042-0.323183\imo$ & $0.007348-0.326434\imo$ \\
 $11$ & $0.000000-0.355670\imo$ & $0.007263-0.357751\imo$ \\
 $12$ & $0.000000-0.387832\imo$ & $0.007189-0.389061\imo$ \\
 $13$ & $0.000091-0.419620\imo$ & $0.007123-0.420364\imo$ \\
 $14$ & $0.000279-0.451184\imo$ & $0.007064-0.451662\imo$ \\
 $15$ & $0.000389-0.482640\imo$ & $0.007011-0.482956\imo$ \\
\hline
\end{tabular}
\caption{Dominant quasinormal modes for the scalar $\ell=0$ perturbations of the Bonanno-Reuter black hole ($M=8$) and the corresponding modes for the Schwarzschild black hole.}\label{table3}
\end{table}

\begin{table*}
\begin{tabular}{c@{\hspace{1em}}|@{\hspace{1em}}c@{\hspace{2em}}c@{\hspace{2em}}c}
\hline
  $\gamma$ & $\omega M$ (WKB) & $\omega M$ (time-domain) & $\omega M$ (accurate) \\
\hline
  $0$     & $0.110678-0.104424\imo$ & $0.110302-0.104679\imo$ & $0.110455-0.104896\imo$ \\
  $4/27$  & $0.111758-0.102964\imo$ & $0.111477-0.103216\imo$ & $0.111600-0.103409\imo$ \\
  $8/27$  & $0.112523-0.101495\imo$ & $0.112587-0.101530\imo$ & $0.112698-0.101699\imo$ \\
  $12/27$ & $0.113241-0.099667\imo$ & $0.113603-0.099566\imo$ & $0.113698-0.099717\imo$ \\
  $16/27$ & $0.114205-0.097505\imo$ & $0.114433-0.097266\imo$ & $0.114512-0.097400\imo$ \\
  $20/27$ & $0.115207-0.094804\imo$ & $0.114902-0.094560\imo$ & $0.114963-0.094678\imo$ \\
  $24/27$ & $0.115156-0.091203\imo$ & $0.114630-0.091441\imo$ & $0.114675-0.091539\imo$ \\
  $28/27$ & $0.112802-0.088751\imo$ & $0.112802-0.088445\imo$ & $0.112827-0.088526\imo$ \\
  $32/27$ & $0.110844-0.087919\imo$ & $0.110664-0.087749\imo$ \\
\hline
\end{tabular}
\caption{Fundamental quasinormal modes for the scalar $\ell=n=0$ perturbations of the black hole.}\label{table4}
\end{table*}

\begin{table*}
\PRDstyle{\begin{tabular}{c@{\hspace{1em}}|@{\hspace{1em}}c@{\hspace{1em}}c@{\hspace{1em}}|@{\hspace{1em}}c@{\hspace{1em}}c}}
\JCAPstyle{\begin{tabular}{c|cc|cc}}
\hline
&\multicolumn{2}{c|\PRDstyle{@{\hspace{1em}}}}{$\ell=0$}&\multicolumn{2}{c}{$\ell=1$}\\
\hline
$n$ & $\omega M$ (Hayward) & $\omega M$ (Schwarzschild) & $\omega M$ (Hayward) & $\omega M$ (Schwarzschild) \\
\hline
  $0$ & $0.113494-0.089160\imo$ & $0.110455-0.104896\imo$ & $0.305627-0.085590\imo$ & $0.292936-0.097660\imo$\\
  $1$ & $0.066731-0.319873\imo$ & $0.086117-0.348053\imo$ & $0.274844-0.263717\imo$ & $0.264449-0.306258\imo$\\
  $2$ & $0.041068-0.576924\imo$ & $0.075742-0.601079\imo$ & $0.218494-0.464862\imo$ & $0.229540-0.540134\imo$\\
  $3$ & $0.021679-0.833067\imo$ & $0.070410-0.853678\imo$ & $0.158106-0.700610\imo$ & $0.203259-0.788298\imo$\\
  $4$ & $0.000000-1.082236\imo$ & $0.067074-1.105630\imo$ & $0.112793-0.957178\imo$ & $0.185109-1.040762\imo$\\
  $5$ & $0.001449-1.317232\imo$ & $0.064742-1.357140\imo$ & $0.081256-1.220106\imo$ & $0.172077-1.294120\imo$\\
\hline
\end{tabular}
\caption{Quasinormal modes for the scalar perturbations of the Hayward black hole ($\gamma=1$) and the corresponding modes for the Schwarzschild black hole.}\label{table5}
\end{table*}

\begin{table}
\begin{tabular}{c@{\hspace{1em}}|@{\hspace{1em}}c@{\hspace{2em}}c}
\hline
  $n$ & $\omega M$ (Hayward) & $\omega M$ (Schwarzschild) \\
\hline
  $0$ & $0.265285-0.080169\imo$ & $0.248264-0.092488\imo$ \\
  $1$ & $0.233370-0.247397\imo$ & $0.214516-0.293668\imo$ \\
  $2$ & $0.172918-0.437335\imo$ & $0.174774-0.525188\imo$ \\
  $3$ & $0.105275-0.663349\imo$ & $0.146177-0.771909\imo$ \\
  $4$ & $0.052269-0.905784\imo$ & $0.126554-1.022551\imo$ \\
  $5$ & $0.015138-1.152384\imo$ & $0.112253-1.273926\imo$ \\
  $6$ &  $0.000-1.441\imo$      & $0.101215-1.525267\imo$ \\
\hline
\end{tabular}
\caption{Dominant quasinormal modes for the electromagnetic perturbations ($\ell=1$) of the Hayward black hole ($\gamma=1$) and the corresponding modes for the Schwarzschild black hole.}\label{table5e}
\end{table}

We know that the algebraically special mode of the Schwarzschild black hole occurs at the 8-th overtone for gravitational perturbation~\cite{Chandrasekhar:1984siy}. On the contrary for both regular black holes considered here, the purely imaginary mode appears in the spectrum of a scalar field. To the best of our knowledge such purely imaginary modes at low overtones is not usual for asymptotically flat black holes, unlike the asymptotically AdS ones (see for instance \cite{Cardoso:2003cj,Konoplya:2017zwo,Gonzalez:2017gwa,Grozdanov:2019uhi} and references therein). One certainly needs to study these purely imaginary modes analytically in order to tell the purely imaginary mode from the mode with a very small real part, and to learn whether they belong to the class of algebraically special modes or not.
Unfortunately, the Frobenius method apparently does not answer this question, because the convergence becomes slow when approaching purely imaginary modes. This is also a reason why we were able to calculate only the first couple of significant digits for such modes. However, we have checked the results with the Bernstein spectral method code \cite{Fortuna:2020obg}, which is particularly useful for finding purely imaginary modes. By comparing with the higher-order expansion in Bernstein polynomials we find that the error of the method is much larger than the usual error for the purely imaginary modes, indicating that the overtones have very small but nonzero real parts.

Another interesting feature of the overtones' behavior can be observed if we compare the real parts of overtones with those for the Schwarzschild case. In the regime when the deviation from the Schwarzschild geometry is tiny, that is, for large mass $M$ for the Bonanno-Reuter metric and small $\gamma$ for the Hayward metric, the fundamental mode naturally differs from the Schwarzschild one insignificantly. However, for the same values of the parameters overtones deviate from their Schwarzschild analogues by hundreds of percent, leaving aside appearance of the purely imaginary modes. We believe that this happens because the overtones are sensitive to the near horizon asymptotic behavior, which is quite different, even once the metrics almost coincide with the Schwarzschild one farther from the horizon. Indeed, it is well-known that the limit $n\to\infty$ is crucially dependent on the event horizon asymptotic \cite{Motl:2003cd}. After all, the overtones are known to be sensitive to boundary conditions in the asymptotic regions as can be seen, for example, for the Schwarzschild-de Sitter black hole \cite{Konoplya:2004uk}.

\section{Conclusions}\label{Conclusions}
Here we considered in detail quasinormal modes of scalar, electromagnetic, and Dirac fields in the background of two types of regular black holes in asymptotically safe gravity: the Bonanno-Reuter and Hayward metrics. The quantum corrections to the geometry of black holes are evidently non-negligible only for sufficiently small Plank-size black holes and can safely be ignored for large astrophysical black holes. Thus, once we talk about relatively small deviation from the Schwarzschild geometry we mean that the fundamental quasinormal mode does not deviate much from its Schwarzschild value, yet, even such small deviation must be in the realm of mini black holes.
Note, that the above considered regular metrics cannot be deduced from the least action principle~\cite{Knorr:2022kqp}.

We have found a number of peculiar features which were missed in the previous studies.
\begin{itemize}
\item The fundamental quasinormal modes are qualitatively similar for both types of identification, that is, the same behavior of the fundamental mode may be appropriate to other ways of identification.
\item The deviation of the fundamental mode from its Schwarzschild value for the Bonanno-Reuter metric can be more than twice larger than it was claimed in the previous studies.
\item Unlike the fundamental mode, overtones are highly sensitive to even small deviation from the Schwarzschild limit, showing deviation of hundreds of percent already at the first few overtones. We believe that this feature may be quite general for various quantum corrected black holes, because the main difference with the classical solution is in the vicinity of the event horizon which is crucial for the overtones' behavior.
\item There are purely imaginary (non-oscillatory) modes in the spectrum of both black holes, which may occur already at the second overtone.
\end{itemize}

In the present paper we have neglected the cosmological term and its dependence on the energy scale. Since the black-hole quasinormal spectrum gets only a small correction due to the $\Lambda$-term~\cite{Zhidenko:2003wq}, such a correction is irrelevant for the Planck-scale black holes, unless one considers a Planck-scale universe, for which it would be difficult to develop a linear perturbation analysis. Although in this case there is another branch of purely imaginary modes associated with the de Sitter universe~\cite{Konoplya:2022xid}, it was shown that this branch of modes does not depend on arbitrary black-hole deformations when the black hole is small compared to the cosmological size~\cite{Konoplya:2022kld}.

Our work could be extended in a number of ways. In particular, the analysis of overtones could be done for other models of quantum corrected black hole metrics which are different from the Schwarzschild limit in the near horizon region, because the same high sensitivity of overtones to even tiny deformations of the geometry is expected in this case. Second, the nature of the purely imaginary mode should be studied: the appealing question is whether this mode is algebraically special or not. Unlike the Schwarzschild algebraically special mode, which has relatively high decay rate and, strictly speaking, does not satisfy the quasinormal boundary conditions, the (almost) non-oscillatory mode of the Hayward black hole corresponds to the second overtone. Therefore, if such a mode appears in the quasinormal spectrum, it manifests itself at the initial stage of the ringdown profile. We hope that further studies will shed light on these questions.

Recently, a solution for the regular Dymnikova-type black-hole spacetime was obtained, which accounts for backreaction effects of the running of the Newton coupling~\cite{Platania:2019kyx}.
The coordinate-independent renormalization group approach has been developed at the level of curvature invariants and the spherically symmetric black-hole metric has been obtained in~\cite{Held:2021vwd}.
These black holes could also be studied in future publications.
Another interesting question would be to learn whether the quasinormal spectra are sensitive to the presence of higher derivatives in the action \cite{Knorr:2019atm,Knorr:2021iwv}.

\begin{acknowledgments}
A.~F.~Z. acknowledges the support from the Silesian University grant SGS/26/2022.
A.~Z. was supported by Conselho Nacional de Desenvolvimento Científico e Tecnológico (CNPq).
A.~Z. thanks Institute of Physics of Silesian University in Opava for the hospitality.
R.~K. would like to acknowledge support of the grant 19-03950S of Czech Science Foundation (GAČR) and Institute of Physics of Oldenburg University for hospitality.
J.~K. would like to gratefully acknowledge support by the DFG Research Training Group 1620 {\sl Models of Gravity} and the DFG project KU 612/18-1.
\end{acknowledgments}

\bibliographystyle{unsrt}
\bibliography{Bibliography}

\end{document}